# X-ray Phase Determination

# by the

# Principle of Minimum Charge

Veit Elser

*Department of Physics*
*Laboratory of Atomic and Solid State Physics*
*Cornell University*
*Ithaca, NY 14853-2501*

When the charge density in a crystal or a quasicrystal is reconstructed from its Fourier modes, the global minimum value of the density is sensitively dependent on the relative phases of the modes. The set of phases that maximizes the value of the global minimum, corresponds, by positivity of the density, to the charge density having the minimum total charge that is consistent with the measured Fourier amplitudes. Phases which minimize the total charge (*i.e.* the average charge density) also have the property that the lowest minima of the charge density become exactly degenerate and proliferate within the unit cell. The large number of degenerate minima have the effect that density maxima are forced to occupy ever smaller regions of the unit cell. Thus by minimizing charge, the atomicity of the charge density is enhanced as well. Charge minimization applied to simulated crystalline and quasicrystalline diffraction data successfully reproduces the correct phases starting from random initial values.

*Dedicated to the memory of Marko V. Jaric*

# 1. Introduction

The electron density in any material can be decomposed into its Fourier modes, the amplitudes of which, $A_\mathbf{q}$, can be measured by x-ray diffraction. If the material is a crystal or even a quasicrystal, the modes $\mathbf{q}$ form a discrete set — a reciprocal lattice — so that a relatively complete portrait of the density is possible in terms of its Fourier amplitudes. Unfortunately, to reconstruct the density one also needs the relative phases of the Fourier modes, $\phi_\mathbf{q}$, which conventional x-ray techniques do not measure. Efforts to solve this "phase problem" either employ additional data (heavy atom substitution, anomalous scattering, three-beam interference) or rely on the principle that only very special sets of phases produce a density that has reasonable physical properties. The latter approach, known as a "direct method", was pioneered by Hauptman and Karle (1953) and applies to the present proposal for solving the phase problem.

Direct methods consider two attributes of the electron density: positivity and atomicity. Since an arbitrary density can be made nonnegative by the addition of a suitable constant, a test of positivity requires an absolute calibration of the diffracted x-ray intensities in terms of the average electron density. This additional burden on the x-ray experiment has favored methods that consider the atomicity of the electron density. The simplest expression of this idea, Sayre's (1952) equation, is a proportionality that exists between the structure factors of the density, $A_\mathbf{q} \exp{(i\phi_\mathbf{q})}$, and their autocorrelation (in the reciprocal lattice), whenever the density consists of well separated identical atoms. Assuming the autocorrelation is dominated by its largest term, Sayre's equation implies approximate relationships between triplets of phases. Generalizations (e.g. unequal atoms, quartet combinations, etc.) and the use of probability theory have led to quite sophisticated algorithms that seek to satisfy relationships among all the phases consistent with a charge density that is highly concentrated at a set of points (atoms) (for a review, see Viterbo, 1992).

A considerably different approach to a direct method solution of the phase problem, described below, grew out of efforts to solve the atomic structure of quasicrystals. The quality of some intermetallic quasicrystals, such as icosahedral AlPdMn, has reached a point where several hundred symmetry inequivalent x-ray reflections can be measured (Boudard *et al.*, 1992). Since a typical icosahedral symmetry orbit has size 120, this implies a wealth of structural data approaching that of protein crystals. Two strategies for reconstructing quasicrystal phases, neither of them direct, have been used in the past. Qiu and Jaric (1990) used the close relationship between an AlCuLi icosahedral phase and a large unit cell crystalline "approximant" — solved by conventional methods — to deduce phases for the quasicrystal. In the case of AlCuFe and AlPdMn icosahedral phases, where convenient "Rosetta stones" such as used by Qiu and Jaric were not available, it was noticed that the intensities of strong reflections behaved in a smooth way with the perpendicular component of the 6-dimensional diffraction vector, $\mathbf{q}_\perp$ (Cornier-Quiquandon *et al.*, 1991; deBoissieu *et al.*, 1994). The vanishing of intensities at certain radii in $\mathbf{q}_\perp$-space then gave a simple rule for sign changes in the assumed centrosymmetric structure factor.

The maximum entropy method, sometimes portrayed as a modern descendent of direct methods (Bricogne, 1984 & 1988), has also been applied to quasicrystals (deBoissieu *et al.*, 1991; Steurer *et al.*, 1993). However, the maximum entropy method is fundamentally a technique for *modelling* the structure (via an arbitrary density) and as such delivers phase information only incidentally. While any attempt at modelling a structure, no matter how crude or specialized, can be viewed as a direct method in that it makes definite predictions about phases, the maximum entropy method is perceived to be sufficiently general to merit special attention. It should be added, however, that whereas the "entropy" in the maximum entropy method has its origin in measurement uncertainty, the phase problem can be addressed even in the case of perfect data. Certainly in this paper it is assumed that the error in the measured amplitudes $A_\mathbf{q}$ is negligible.

The difficulty in adapting conventional direct methods to quasicrystals stems from the very different nature of the charge density in the "unit cell". The example of a 1-dimensional quasicrystal (or incommensurately modulated structure) having *two* fundamental periods in its reciprocal lattice already illustrates this difference. As shown in figure 1, the quasiperiodic 1-dimensional density is obtained as a cut through a periodic density in a 2-dimensional space. Atoms, or point-like concentrations of charge on the cut, become extended curves in the 2-dimensional unit cell. In general, atomic charge distributions acquire additional dimensions corresponding to the extra periods in the reciprocal lattice. In icosahedral quasicrystals, for example, where the reciprocal lattice has 6 periods, atoms are 3-dimensional "surfaces". Although still "compact" in the sense of having a lower dimensionality than the unit cell, the charge distributions in quasicrystals are clearly at odds with the logic that lead to direct methods in the first place. For point-like charge distributions (as in crystals) it was argued that a *finite* number of parameters completely specify the density; an overabundance of measured diffraction amplitudes then leads to an overdetermined set of equations and the theoretical possibility of deducing phases. In contrast, to specify just one atomic surface in a quasicrystal, such as the curve shown in figure 1, already an infinite number of parameters is required.

The more primitive property of positivity makes no distinction between the kinds of charge distributions in crystals and quasicrystals. A phase determination principle formulated only in terms of positivity should therefore apply to crystals and quasicrystals alike. Such a principle, apparently not considered before, might be called "the principle of minimum charge". Specifically: of all the charge densities that can be realized by every possible choice of phases, the unique "atomic" charge density corresponds to the phase choice that *minimizes the average*

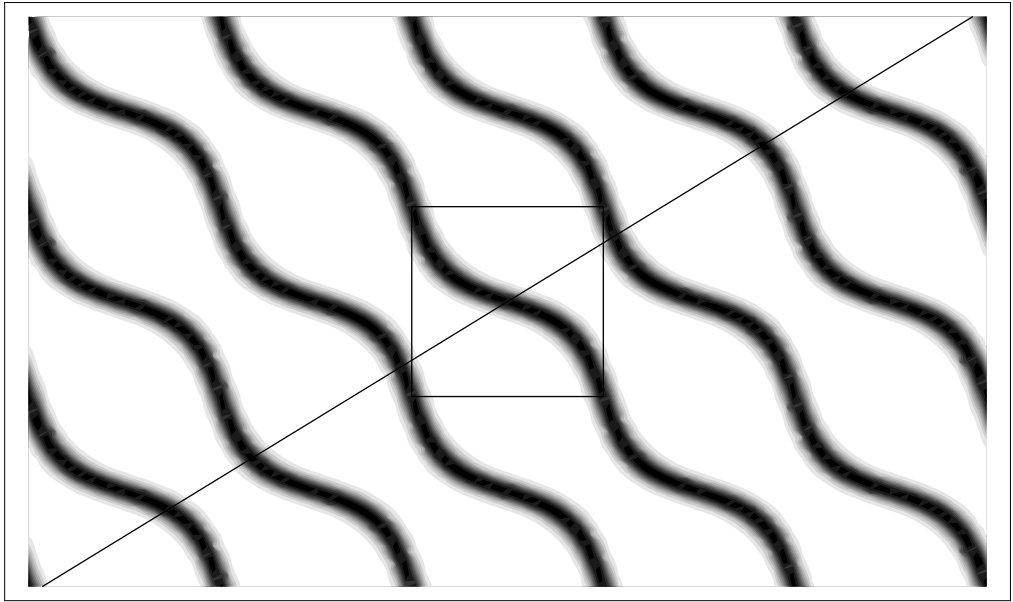

Figure 1. One-dimensional quasiperiodic density obtained as a cut through a 2-dimensional density having the periodicity of the square lattice. Infinitely many parameters are in principle required to specify even one "atomic surface".

*charge density consistent with the demands of positivity.* As discussed in greater detail below (section 2), the value of the average charge density is determined by the requirement that the value of the density at its global minimum in the unit cell takes the smallest possible value: zero. An important property of an "atomic" density, as determined by the principle of minimum charge, is the fact that the global minimum of the density is not unique but is highly degenerate. Expressed in more qualitative terms, the minimum charge principle inevitably finds phases that generate large expanses of nearly constant density modulated by tiny ripples. Whereas the ripples are an artifact of the truncation of the Fourier series, they play a key role in a reasonably successful algorithm for finding states of minimum charge (section 3). Application of this algorithm to simulated crystalline and quasicrystalline charge densities in 2-dimensions (section 4) gives promising results and opens the possibility that the principle of minimum charge may be developed into a practical direct method of phase determination.

## 2. The principle of minimum charge

In what follows we make no distinction between crystals and quasicrystals. The charge density is in both cases given by the Fourier series

$$\rho(\mathbf{x}) = \rho_0 + 2\sum_{\mathbf{q}} A_{\mathbf{q}} \cos(\mathbf{q} \cdot \mathbf{x} - \phi_{\mathbf{q}}) , \tag{1}$$

where the terms in the sum involve one representative from each pair $\{\mathbf{q}, -\mathbf{q}\}$ of nonzero reciprocal lattice vectors; $\rho_0 > 0$ is the average charge density, $A_{\mathbf{q}}$ the known (positive) amplitudes, and $\phi_{\mathbf{q}}$ the unknown phases. The dimensionalities of $\mathbf{q}$ and $\mathbf{x}$ are equal to the number of periods of the reciprocal lattice, known as the rank (Rokhsar *et al.*, 1987). In the case of quasicrystals the physical density (in a lower dimension) is obtained from $\rho(\mathbf{x})$ by the construction shown in figure 1. To obtain a finite Fourier series we exclude terms whose amplitudes are less than some multiple $\eta < 1$ of the maximum amplitude. A typical value of $\eta$, or "truncation parameter", used in our simulations (section 4) is $0.1$. To emphasize the finiteness of our Fourier series we will sometimes use the notation $\mathbf{q} = 1, \ldots, N$ when the series has $N$ terms (in some arbitrary order).

Since the average charge density $\rho_0$ is usually not known, and in any event measured by completely different means than the amplitudes $A_{\mathbf{q}}$, we define a reduced charge density given by

$$\tilde{\rho}(\mathbf{x}) \equiv \rho(\mathbf{x}) - \rho_0 . \tag{2}$$

Because the integral of $\tilde{\rho}(\mathbf{x})$ over the unit cell vanishes, the value of $\tilde{\rho}$ at its global minimum, $\tilde{\rho}(\mathbf{x}_{\min})$, is negative. Given some $\tilde{\rho}(\mathbf{x})$, that is, a particular set of phases, the minimum value of $\rho_0$ consistent with positivity of $\rho(\mathbf{x}_{\min})$ is

$$\rho_0 = -\tilde{\rho}(\mathbf{x}_{\min}) > 0 . \tag{3}$$

Expressed in these terms, the principle of minimum charge corresponds to finding phases that minimize $\rho_0$, or equivalently, maximize the global minimum value, $\tilde{\rho}(\mathbf{x}_{\min})$, of the reduced charge density. Henceforth we let equation (3) define the average charge density.

A 1-dimensional example serves to illustrate how the principle of minimum charge selects densities with atomic characteristics. Shown in figure 2(a) is the reduced charge density com-puted from the diffraction amplitudes and phases of a unit cell of three atoms having identical Gaussian charge distributions. The wiggles in the background are due to the truncation of the Fourier series. A qualitatively different reduced density, shown in figure 2(b), is obtained when the phases are assigned random values. While neither of the reduced densities is positive, the statistical symmetry between positive and negative values in the case of random phases is clearly absent when the phases have their correct values. The connection with the principle of minimum charge is seen by noticing that a much larger average density, $\rho_0$, must be added to the reduced density of figure 2(b) to arrive at a nonnegative total density.

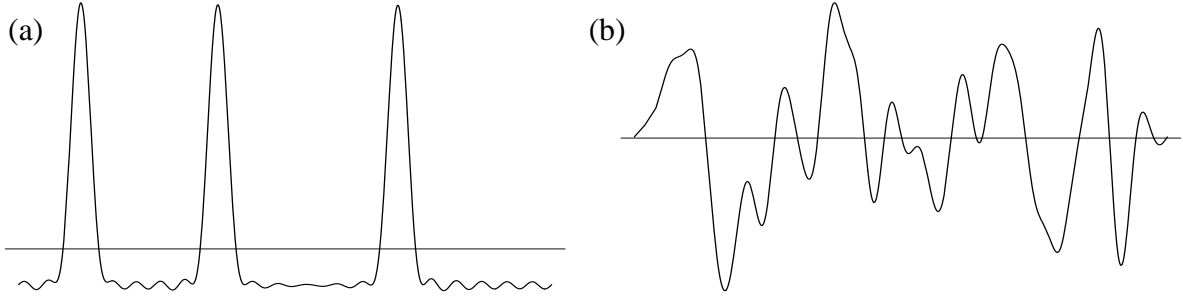

Figure 2. Reduced density reconstructed from the correct phases (a), and random phases (b).

The low value of the global minimum value of $\tilde{\rho}(\mathbf{x})$ in figure 2(b) is easily improved upon, while the global minimum in figure 2(a) is in a "competition" with several local minima having only slightly higher values. Figure 3(a) gives a magnified view of figure 2(a), showing the competition among local minima. Because of truncation, the correct phases in this example do not realize the minimum possible value of $\rho_0$. The true minimum, *i.e.* the "optimal" density, was found using the algorithm described in section 3 and has a value of $\rho_0 = \rho_{opt}$ that is reduced by 4%; the resulting reduced density is shown in figure 3(b). Although insignificant from a physical point of view, the difference between figures 3(a) and (b) illustrates the basic mathematical property that distinguishes an optimal density: *the global minimum has become multiply degenerate*. This basic fact is seen most directly by linearizing $\tilde{\rho}(\mathbf{x}_{min})$ with respect to the phases near an optimal set of phases. Since this forms the basis of our algorithm for finding optimal phases, we defer the discussion of this point to section 3. Here we let this fact be the inspiration for more sweeping (and unproven) assumptions that allow us to relate charge minimization to the enhancement of atomicity.

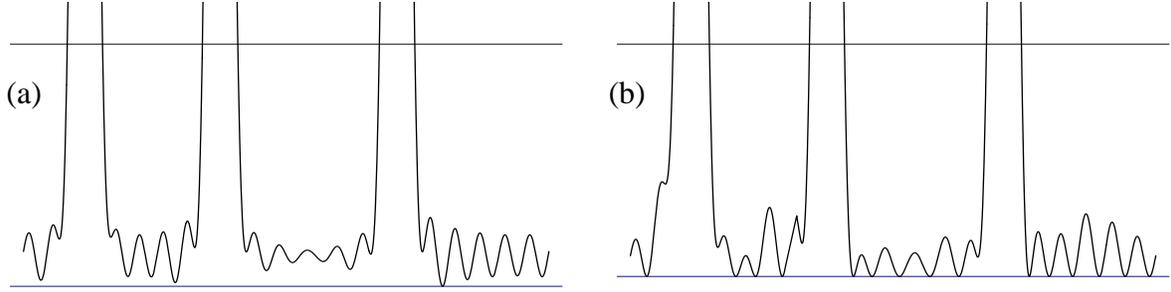

Figure 3. A slight adjustment of the phases in (a), which maximize the density at the global minimum, produces the density shown in (b) where the global minimum has become multiply degenerate.

The following discussion focuses on the distribution of reduced density values in the unit cell. Given some reduced density $\tilde{\rho}(\mathbf{x})$, the normalized distribution function is defined by

$$F(\rho;\tilde{\rho}(\mathbf{x})) = \int \delta(\tilde{\rho}(\mathbf{x}) - \rho) \frac{d\mathbf{x}}{V} \quad , \tag{4}$$

where $\delta(\ )$ is the Dirac delta function, and $d\mathbf{x}$ and $V$ are respectively the volume element and volume of the unit cell. An average distribution function is obtained by averaging the distributions (4) over the subensemble of reduced densities having a fixed value of average charge, $\rho_0$:

$$F(\rho;\rho_0) = \frac{\int F(\rho;\tilde{\rho}(\mathbf{x}))\delta(\tilde{\rho}(\mathbf{x}_{min}) + \rho_0)\frac{d\phi_1}{2\pi}\cdots\frac{d\phi_N}{2\pi}}{\int \delta(\tilde{\rho}(\mathbf{x}_{min}) + \rho_0)\frac{d\phi_1}{2\pi}\cdots\frac{d\phi_N}{2\pi}} \quad . \tag{5}$$

Here we have assumed that the Fourier series (1) is truncated after $N$ terms so that $\tilde{\rho}(\mathbf{x})$ and its minimum value, $\tilde{\rho}(\mathbf{x}_{min})$, are functions of $N$ phases. The earlier remarks about the multiplicity of the global minimum in an optimal density, together with the results of numerical experiments with charge minimization (section 4), suggest that the average distribution function (5) approaches the L-shaped form shown in figure 4 as $\rho_0$ tends toward its optimal (minimum) value $\rho_{opt}$.

Borrowing the language of Bose-Einstein condensation, the distribution of figure 4 should be viewed as the sum of two distributions: a sharply peaked "condensate" near $-\rho_0$ and a broad, relatively smooth distribution. For the latter we will assume a one-parameter scaling form, the single parameter being simply its first moment, $\bar{\rho}$. In detail, our main assumption is the statement

$$\lim_{\rho_0 \to \rho_{opt}} F(\rho;\rho_0) \sim f_0 \delta(\rho + \rho_0) + (1 - f_0)(1/\bar{\rho})\hat{F}(\rho/\bar{\rho}) \quad , \tag{6}$$

where $\hat{F}$ is the (normalized) smooth distribution and $f_0$ is a number between 0 and 1 representing the relative weight of the condensate peak. In physical terms $f_0$ represents the fraction of the unit cell having vanishingly small charge density. We note that the delta-function in (6) will be broadened by truncation effects.

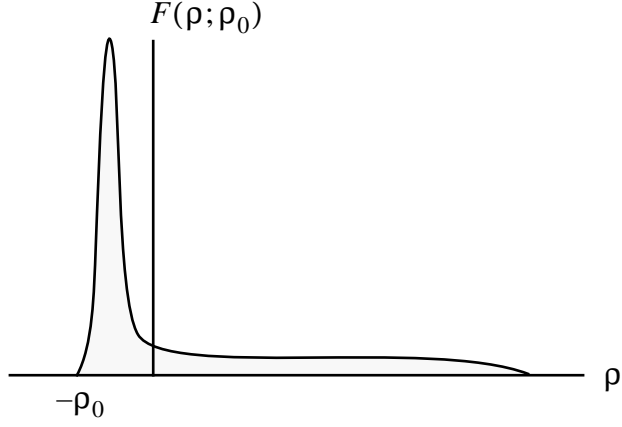

Figure 4. Form of the average density distribution function when the average charge density, $\rho_0$, is near its minimum (optimal) value.

From (6) it is straightforward to make the connection between charge minimization and atomicity. Since $\bar{\rho}$ is defined to be the first moment of the smooth distribution we have

$$\int \hat{F}(y) y \, dy = 1 \, . \tag{7}$$

The second moment defines a numerical constant $\sigma$,

$$\int \hat{F}(y) y^2 \, dy = \sigma > 1 , \qquad (8)$$

where the inequality follows from the fact that $\hat{F}$ has a finite variance. Since $F(\rho;\rho_0)$ is a distribution function for *reduced* densities, its first moment vanishes. Consequently,

$$-\rho_0 f_0 + \bar{\rho}(1 - f_0) = 0 . \qquad (9)$$

Returning for a moment to the particular distribution function $F(\rho;\tilde{\rho}(\mathbf{x}))$, we note that its second moment is given directly in terms of the Fourier amplitudes:

$$\int F(\rho;\tilde{\rho}(\mathbf{x}))\rho^2 d\rho = \int \tilde{\rho}^2(\mathbf{x}) \frac{d\mathbf{x}}{V} = 2\sum_{\mathbf{q}} A_{\mathbf{q}}^2 \equiv I . \qquad (10)$$

Because (10) holds for every member of the subensemble used in the definition of the average distribution function, the same second moment condition holds for $F(\rho;\rho_0)$. Together with equation (8) we then obtain

$$\rho_0^2 f_0 + \sigma \bar{\rho}^2 (1 - f_0) = I . \qquad (11)$$

Eliminating $\bar{\rho}$ between the first and second moment relations (equations (9) and (11)) we arrive at our final result:

$$\frac{\rho_0^2}{I} = \frac{1 - f_0}{f_0(1 + (\sigma - 1)f_0)} . \qquad (12)$$

Equation (12) expresses $\rho_0$ as a monotonically decreasing function of $f_0$ (see figure 5). As $\rho_0$ decreases, the "non-condensate" fraction of the unit cell volume, $1 - f_0$, must also decrease, *i.e.* the charge density becomes more concentrated or "atomic". For a reasonably atomic density ($f_0 \cong 1$) the fractional volume occupied by atoms (or "atomic surfaces" in the

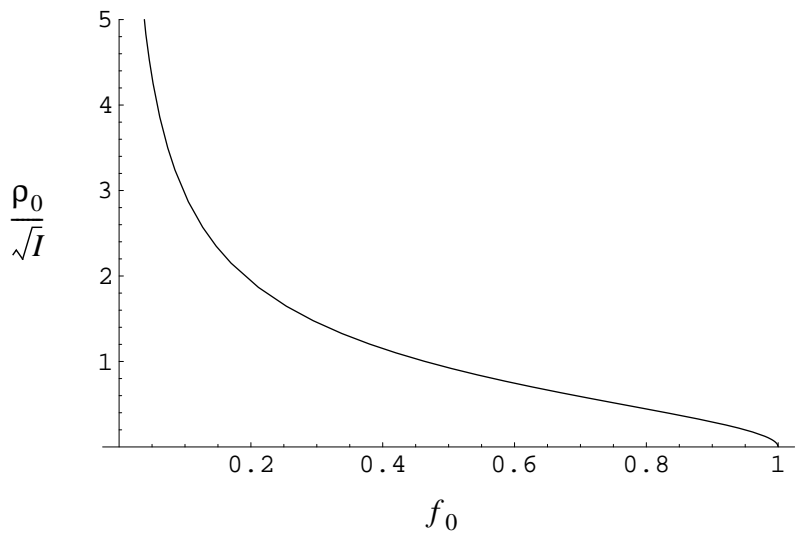

Figure 5. Relationship between positivity and atomicity as given by equation (12) (with $\sigma = 4/3$): as the fractional volume *not* occupied by charge, $f_0$, increases, the average charge density $\rho_0$ must decrease.

case of quasicrystals) is well approximated by

$$1 - f_0 \cong \sigma \frac{\rho_0^2}{I}. \tag{13}$$

One outcome of this analysis is that the dimensionless expression $\rho_0/\sqrt{I}$ emerges as a very natural figure of merit in phase determination. In addition to having a very clear interpretation, it has the advantage over other commonly used figures of merit (Viterbo, 1992) in being directly proportional to what is being optimized.

We conclude this section with the observation that there are infinitely many examples of charge densities where the method of charge minimization is guaranteed to fail. A typical example in one dimension is the charge density of figure 2(b). The Fourier waves of this hypothetical (and exotic) periodic material, when combined with somewhat different phases, produces the much more atomic density shown in figure 2(a). The principle of charge minimization *always* returns an atomic density, even if the material generating the x-ray data has very nonatomic characteristics. To bring this potentiality for "failure" into focus, consider a crystal of atoms having a freely variable size. As the atomic size reaches the interatomic scale, and the corresponding charge distributions begin to overlap, the interstitial region — best charac-terized as having an essentially constant density value — shrinks in volume. Since the minimum charge principle selects distributions with a significant interstitial volume (or "condensate" fraction $f_0$), there comes a point when the atoms in this hypothetical situation are so large that the application of charge minimization would produce a completely different, albeit atomic, density. To avoid the possibility of such "imposters" it is not always necessary, however, for the x-ray data to extend to atomic scale resolution. For example, in complex molecular crystals, where atomic resolution is rarely achieved, there will typically be large constant-density voids even when individual atoms cannot be resolved.

# 3. An algorithm for minimizing charge

Given two sets of phases, the principle of minimum charge provides a simple criterion for selecting the more "atomic" of the two: find the global minima of the corresponding reduced charge densities and pick the one having the largest value. In general, however, atomicity is more reliably evaluated by the application of specialized knowledge so that the primary motivation for charge minimization is the ease of its implementation in an automated procedure for phase determination. Charge minimization even goes to extremes in avoiding specialized knowledge, in particular, by not making a distinction between crystalline and quasicrystalline charge densities.

The global minimum value of a function is just the minimum of its values at its local minima. This forms the basis of an iterative search procedure for phases, where each iteration step is a problem in linear programming. For any particular set of phases,

$$\phi_{\mathbf{q}}, \qquad \mathbf{q} = 1, ..., N, \tag{14}$$

we let

$$\mathbf{x}_{\mathbf{p}}, \qquad \mathbf{p} = 1, ..., M \tag{15}$$

denote the set of local minima of the reduced density $\tilde{\rho}(\mathbf{x})$. This set is finite because the Fourier series for $\tilde{\rho}(\mathbf{x})$ is finite ($N$ terms). Now if $\rho_{\min}$ is the value of the global minimum, then

$$\rho_{\min} \leq \tilde{\rho}(\mathbf{x}_{\mathbf{p}}), \qquad \mathbf{p} = 1, ..., M. \tag{16}$$

Here the phases appear explicitly in the definition of $\tilde{\rho}(\mathbf{x})$ and implicitly via the positions of the local minima, $\mathbf{x}_{\mathbf{p}}$. In fact, even the *number* of local minima, $M$, is a function of the phases.

By restricting the phases to lie in a small interval about a set of starting values,

$$-\Delta \leq \phi_{\mathbf{q}} - \phi_{\mathbf{q}}(0) \leq \Delta, \qquad \mathbf{q} = 1, ..., N, \tag{17}$$

we can approximate (16) by a set of linear inequalities for the phases,

$$\rho_{\min} \leq \tilde{\rho}(\mathbf{x}_{\mathbf{p}}(0)) + \sum_{\mathbf{q}=1}^{N} g_{\mathbf{pq}}(\phi_{\mathbf{q}} - \phi_{\mathbf{q}}(0)), \qquad \mathbf{p} = 1, ..., M, \tag{18}$$

where $\mathbf{x_p}(0)$ are the local minima corresponding to the starting set of phases and $g_{\mathbf{pq}}$ is the $M \times N$ matrix of coefficients

$$g_{\mathbf{pq}} = \frac{\partial}{\partial \phi_{\mathbf{q}}} \tilde{\rho}(\mathbf{x_p})\bigg|_{\phi_{\mathbf{q}} = \phi_{\mathbf{q}}(0)}$$
$$= 2A_{\mathbf{q}} \sin(\mathbf{q} \cdot \mathbf{x_p}(0) - \phi_{\mathbf{q}}(0)) \,. \tag{19}$$

Inequalities (17) and (18) represent a linear programming problem for the set of $N+1$ variables

$$\{\rho_{min}, \phi_1, ..., \phi_N\}\,, \tag{20}$$

with objective function (to be maximized) given by $\rho_{min}$. Clearly this system always has a "feasible point" so a solution, denoted by

$$\{\rho'_{min}, \phi_1(1), ..., \phi_N(1)\}\,, \tag{21}$$

always exists. Finally, our phase determination algorithm consists in iterating the mapping given by

$$\{\phi_1(0), ..., \phi_N(0)\} \rightarrow \{\phi_1(1), ..., \phi_N(1)\} \,. \tag{22}$$

If we regard (22) as a discrete dynamical system, then solutions of the phase problem would appear to be the corresponding attractive fixed points. Provided $\Delta$ in (17) is small enough, the neglect of nonlinear terms in (18) is a good approximation and the value of $\rho_{min}$ should increase with each iteration, approaching some limiting value at the fixed point. We recall that the maximization of $\rho_{min} = \tilde{\rho}(\mathbf{x}_{min})$ corresponds, by (3), to a minimization of the average charge density.

Due to the very different nature of the constraint equations (17) and (18), however, the situation is not quite this simple. Numerical experiments (section 4) show that the number of local minima, $M$, is typically smaller than the number of variables, $N+1$. As a result, the solution point in the linear programming problem is typically determined by a significant subset of the $2N$ constraints (17) in addition to the $M$ constraints in (18). In other words, in every iteration some subset of the phases will be determined to lie on the boundary of the current hypercubic domain (17).

To discuss the dynamics of the algorithm it helps to have a geometrical picture of the function being maximized: $\tilde{\rho}(\mathbf{x}_{min})$, where $\mathbf{x}_{min}$ is the global minimum of $\tilde{\rho}$ in the unit cell. Since the local minima of $\tilde{\rho}$ are locally smooth functions of the phases, the condition that some $M$ local minima are exactly degenerate defines a submanifold of codimension $M-1$ in the $N$ dimensional torus of phases. On each such submanifold $\tilde{\rho}(\mathbf{x}_{min})$ is a smooth function. A useful comparison is provided by linear programming, where the submanifolds correspond to the faces of a convex polytope and the objective function is linear. While this structure applies locally to the function $\tilde{\rho}(\mathbf{x}_{min})$, it should not be overlooked that the true submanifolds are curved and the objective function is nonlinear. It is for this reason that the linear programming step of the algorithm must be iterated.

The dynamical progress of the algorithm toward a solution depends critically on the value of its single parameter, $\Delta$. In the limit of small $\Delta$ the linear approximation becomes exact and the linear programming iterates will lie close to submanifolds where a certain number of minima (of $\tilde{\rho}$) are exactly degenerate. This has the drawback that many iterations are required to reach a solution, and more seriously, progress toward a solution may halt if $\tilde{\rho}(\mathbf{x}_{min})$ has a local maximum in the interior of some submanifold which is not the global maximum. Numerical experiments described in the next section show that the latter does indeed occur so that a small value of $\Delta$ can be used only if the phases are already known to be near a solution. In the appendix we show that the charge minimization problem is closely related to a hard combinatorial optimization problem, so that the possibility of "false" maxima is hardly surprising. For phase determination with little or no *a priori* information, it is therefore necessary to use values of $\Delta$ large enough to avoid the problem of false maxima. Although still deterministic, the behavior of the algorithm in this regime can be characterized as an example of biased diffusion: the increased strength of nonlinearities at large $\Delta$ correspond to a random force, while each quasi-random step is clearly biased in favor of maximizing $\tilde{\rho}(\mathbf{x}_{min})$. Remarkably, this diffusion process is empirically quite successful in finding true solutions. Numerical experiments show that a value $\Delta \approx 0.5$ is neither too small to run the risk of being trapped, nor too large to invalidate the linear approximation that forms the basis of the algorithm.

The computationally most intensive step in each iteration is to find the locations of the local minima of $\tilde{\rho}$. Since the number of local minima scales as the number of atoms in the unit cell, while the number of terms in the expression for $\tilde{\rho}$, at fixed resolution, also scales as the number of atoms, the computational cost at each iteration scales as the square of the number of atoms. To estimate the number of iterations needed to find a solution we distinguish between *ab initio phase determination,* where all phases are initially unknown, and *phase refinement,* where a relatively small number $N_s$ of phases, corresponding to large amplitudes $A_\mathbf{q}$ ("strong reflections"), are assumed known. The phases of the strong reflections locate the basin of attraction that the diffusion process must find in the case of *ab initio* phase determination. Assuming the extent of this basin is roughly $\pi$ in each of $N_s$ dimensions, the diffusion process can expect to take some $2^{N_s}$ steps to find the basin. Once the basin is found, we expect the bias in the random walk to be strong and the remaining number of iterations to be relatively independent of the number of phases (atoms). In summary, we expect the computational cost of phase refinement to scale as the square of the number of atoms, while for *ab initio* phase determination this cost must be multiplied by a factor of order $2^{N_s}$, where $N_s$ is the number of strong reflections. A theoretical model of how $N_s$ should scale with the number of atoms is presently lacking, but an overall exponential growth in computational cost with the number of atoms seems appropriate, given the close relationship to the knapsack optimization problem (see appendix).

# 4. Examples of phase determination in two dimensions

Motivated by the ease of visualizing charge densities in two dimensions, it was decided to perform phase determination on simulated crystalline (point-like) and quasicrystalline (curvilinear) densities in two dimensions. Since our main interest was the behavior of the charge minimization algorithm in simple situations, there was no effort to optimize the implementation of the algorithm. The results presented here were obtained on a 120 MHz Macintosh running an approximately 30 line program written in the *Mathematica* language.

Features common to all of our simulations are a square Bravais lattice with unit lattice parameter, a total charge normalized to unity, and a truncation parameter $\eta = 0.1$. All examples, even the centrosymmetric ones, were treated without regard to point group symmetry. The solutions therefore form two families: translates of the true density, and translates of the inverted true density. To evaluate the progress of the phase determination we therefore computed two overlap functions:

$$Q_\pm = \max_{\mathbf{y}} \left\{ \frac{\int \tilde{\rho}(\mathbf{x}) \tilde{\rho}^{\pm T}(\mathbf{x}+\mathbf{y}) d\mathbf{x}}{\int \tilde{\rho}^2(\mathbf{x}) d\mathbf{x}} \right\} = \max_{\mathbf{y}} \left\{ \frac{\sum_{\mathbf{q}} A_{\mathbf{q}}^2 \cos(\mathbf{q}\cdot\mathbf{y} + \phi_{\mathbf{q}} \mp \phi^T{}_{\mathbf{q}})}{\sum_{\mathbf{q}} A_{\mathbf{q}}^2} \right\}, \tag{23}$$

where $\tilde{\rho}^{\pm T}$ are the true densities given by the two inversion related sets of true phases, $\phi^T{}_{\mathbf{q}}$. A solution thus corresponds to either $Q_+$ or $Q_-$ having the value unity (or both, in the case of centrosymmetry).

Crystalline charge densities were, for simplicity, taken to be randomly placed, identical atoms with Gaussian form factors. Explicitly, the structure factors are

$$A_{\mathbf{q}} \exp(i\phi_{\mathbf{q}}) = \left[ \frac{1}{N} \sum_{k=1}^{N} \exp(i\mathbf{q}\cdot\mathbf{x}_k) \right] \exp(-B\mathbf{q}\cdot\mathbf{q}), \tag{24}$$

where $\mathbf{x}_1, \mathbf{x}_2, \ldots$ are the randomly chosen positions. Since the dimensions of the unit cell are fixed, we set $B = 0.3/N$ to ensure that the atomic size scales correctly with the mean atomic separation. To further improve the atomicity of the density we removed atoms if their separation from another atom was less than $0.6/\sqrt{N}$.

Figure 6 shows an example of *ab initio* phase determination for 10 atoms, a situation where $\eta = 0.1$ gives a Fourier series of 92 terms. The first frame, fig. 6(a), shows the initial density corresponding to random values assigned to all 92 phases. Subsequent frames show the evolution of the density after 100, 200, and 360 iterations of the charge minimization algorithm of section 3. The variation of the overlaps $Q_\pm$ are shown in figure 7. During the first 200 iterations the maximum phase angle change per iteration was set at $\Delta = 0.5$. Already after just a few iterations the density becomes concentrated at points, initially few in number. During the course of the minimization atoms appear and disappear throughout the unit cell. After 100 iterations (fig. 6(b)), for example, we note that some of the well established charge concentrations correspond to correct atomic positions (fig. 6(d)), while others are clearly imposters. As the number of cor-rect positions grows, we reach a point after 200 iterations (fig. 6(c)) where the value of $Q_-$ is close to unity and a solution has been found. To identify this point in the minimization without the benefit of knowing the true phases (and hence $Q_\pm$) is not difficult: the subsequent evolution of the charge density only exhibits small fluctuations in the atomic positions and an overall drift in the origin. Fluctuations are reduced and the phases determined to higher accuracy by decreasing the parameter $\Delta$. In the example shown, $\Delta$ was reduced to 0.2 at iteration 200, and then further to 0.1 at iteration 300. The final density (fig. 6(d)), at iteration 360, shows essentially the same atomic structure found with $\Delta = 0.5$ (fig. 6(c)), but with more circular atoms and a flatter background.

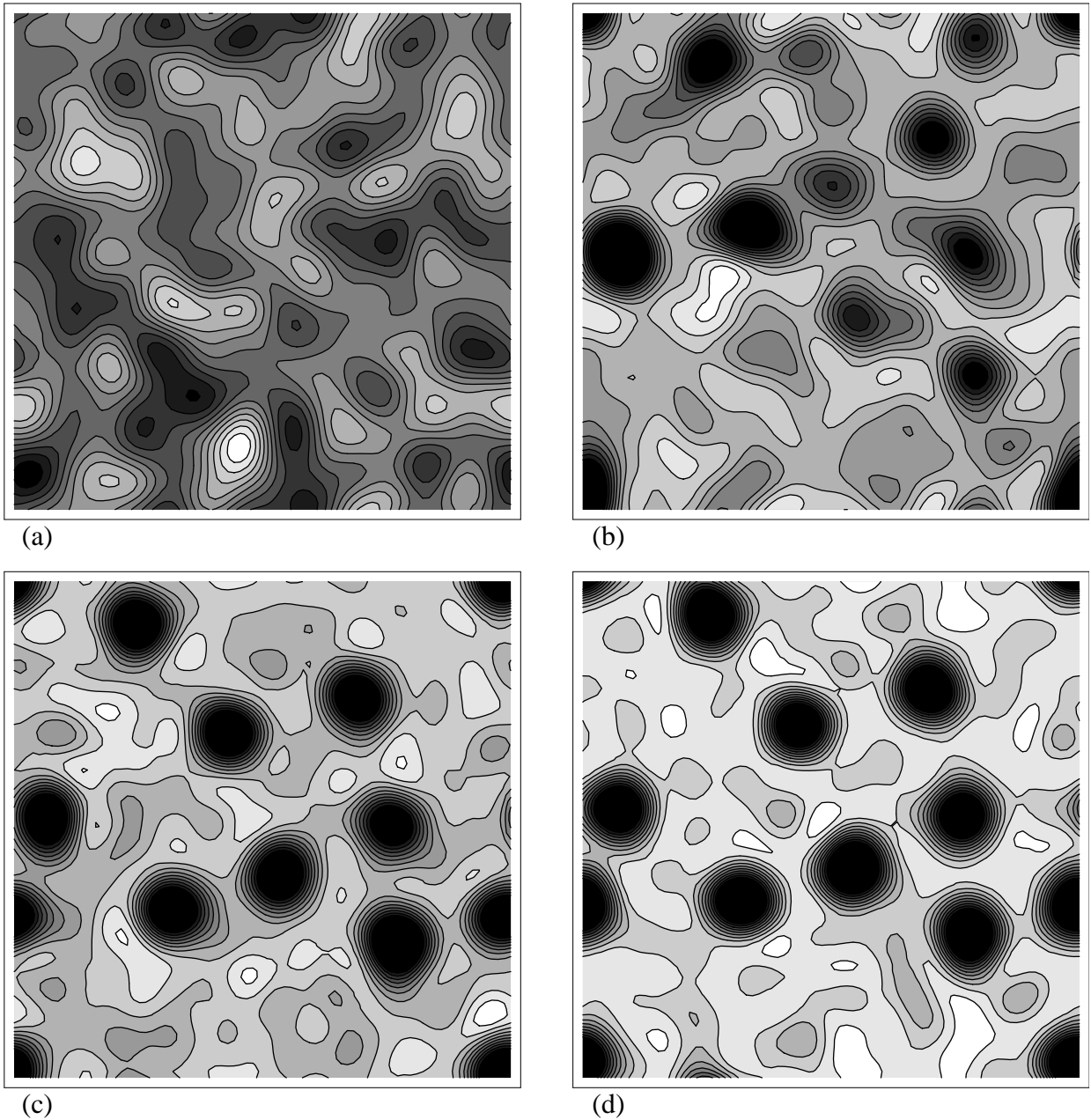

Figure 6. Evolution of the density in a crystal with 10 equal atoms per unit cell. Frames (b)-(d) show the results of respectively 100, 200, and 360 iterations of the charge minimization algorithm applied to the intial random state (a).

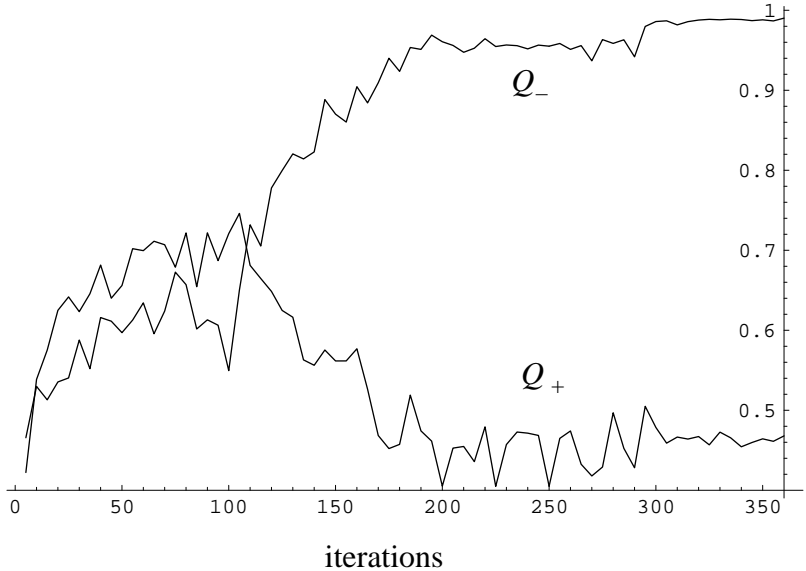

Figure 7. Corresponding evolution of the two overlaps, $Q_\pm$, for the charge minimization example of figure 6.

The bias in the diffusive dynamics of the phases originates from data computed at the $M$ local minima of the reduced density (equation (18)). A simple measure of the strength of the bias is simply to compare $M$ with the number of unknown phases, or $N = 92$ in the present example. Following the behavior observed in all other experiments, $M$ increased with decreasing $\Delta$ but was always smaller than $N$, even for $\Delta = 0.1$, where $M$ was typically 50. This means that some fraction of the phases make steps of size exactly $\Delta$ with every iteration. When the true density has been found, however, such phases will tend to make compensating steps of $+\Delta$ and $-\Delta$ in successive iterations.

If the initial value of $\Delta$ is too small, the algorithm invariably gets trapped in a metastable state. Starting from the same random assignment of phases as in fig. 6(a), the result of 100 iter-ations with $\Delta = 0.1$ is shown in fig. 8(a). Although this density persists with further iterations, it can easily be distinguished from the correct density, fig. 6(c), which persists at $\Delta = 0.5$. The simplest procedure is to check for reproducibility (up to translation and inversion) of the final density for different random initial phases. A repeat of the previous experiment, for example, results in the completely different density shown in fig. 8(b). Also, given a set of candidate densities, say figs. 6(d), 8(a), and 8(b), one can compare their global minimum values since that is, after all, what these densities are trying to maximize. For the three examples cited, these are respectively $\tilde{\rho}(\mathbf{x}_{min}) = -1.72$, -2.38, and -2.41 and show that 6(d) is clearly more "atomic" than either 8(a) or 8(b). The density values at the local minima form strongly peaked distributions. The mean values of the distributions corresponding to figs. 6(d), 8(a), and 8(b) are given by -1.38, -2.03, and -1.83 and provide a somewhat more meaningful comparison than the least element in each distribution. The distribution of *all* the density values in the unit cell are compared in figure 9. We note that the true density (corresponding to fig. 6(d)) comes closest to the L-shaped distribution introduced in section 2 (fig. 4).

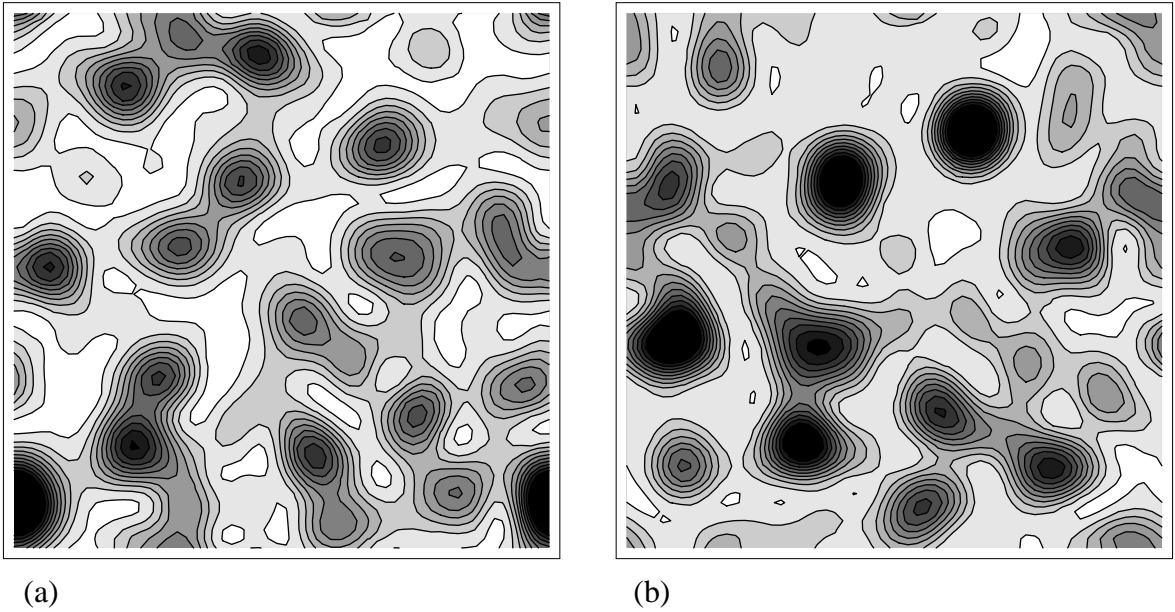

Figure 8. Two final states of charge minimization when the parameter $\Delta$ is too small. Densities (a) and (b) are the result of 100 iterations on the same 10 atom system shown in fig. 6, starting with different random initial phases.

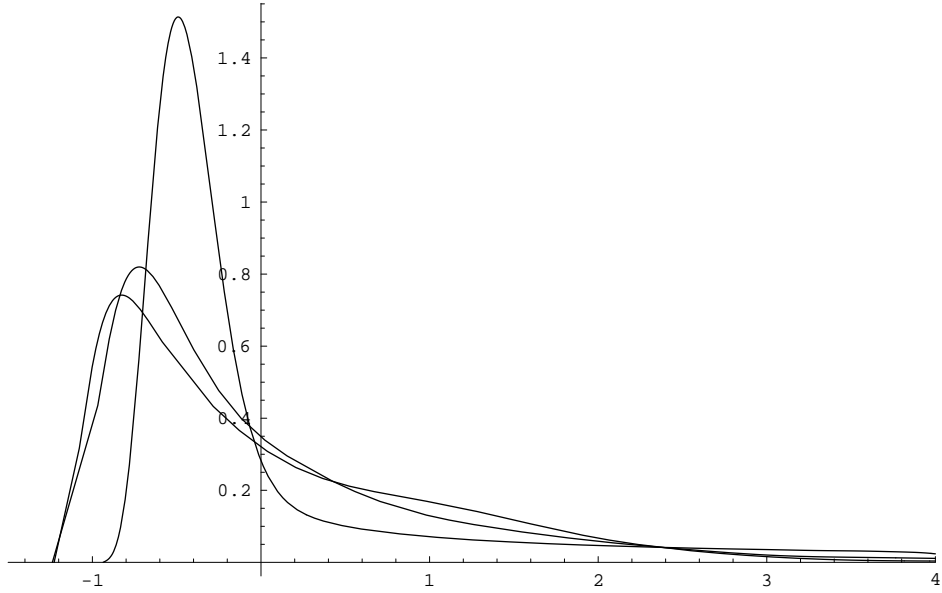

Figure 9. Density distributions in the unit cell corresponding to (in order of decreasing peak height) figs. 6(d), 8(b), and 8(a).

As a simple example of a quasicrystalline density we chose the continuous atomic surface with centrosymmetric modulation shown in figure 10(a). The structure factor has the form

$$A_{\mathbf{q}} \exp(i\phi_{\mathbf{q}}) = J_{m+n}(\gamma(m-n)) \exp(-B(m-n)^2) \ , \qquad (25)$$

where $\mathbf{q} = 2\pi(m,n)$, $J$ is the Bessel function, $\gamma$ controls the amplitude of the modulation, and $B$ the thickness of the atomic surface. In the example shown, $\gamma = 0.5$ and $B = 0.01$; the resulting truncated Fourier series had 44 terms.

As with the crystalline example, the initial phases were given random values and $\Delta$ was set to the value 0.5. The density evolution is portrayed in figures 10(b-d) and shows a rapid convergence to a translate of the true density. A plot of the overlap function $Q_+$ ($Q_-$ being identical) is shown in figure 11. The final density proved to be quite stable; upon further minimization (not shown) the curvilinear density never showed a tendency to break up into pointlike atoms, for example.

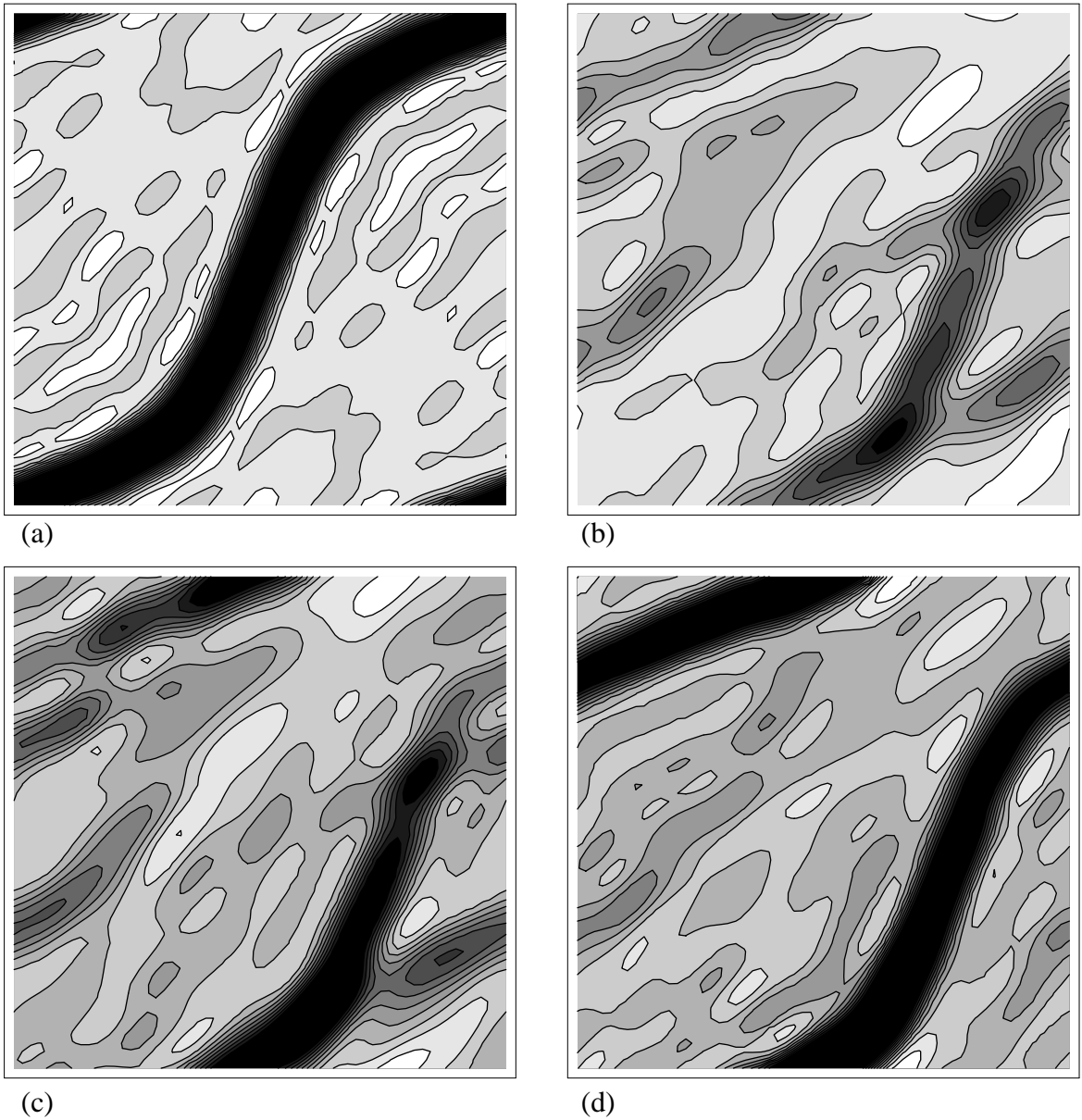

Figure 10. Example of phase reconstruction for a curvilinear charge density as in a quasicrystalline atomic surface. Frame (a) shows the true density while (b)-(d) are the results of respectively 10, 40, and 80 iterations of the charge minimization algorithm.

To investigate the performance of the charge minimizing algorithm in phase refinement, a random system of 20 atoms (178 Fourier waves) was studied with two kinds of initial conditions. In the case of "uniform randomization", a random phase error uniformly distributed within $\pm\delta\pi$ ($0 < \delta < 1$) was added to all the true phases. The other case, "selective randomization", was similar but used $\delta = 0.5$ on only the $N_s$ largest amplitude waves and completely randomized the rest ($\delta = 1$). The criterion for evaluating success was also different from that used in the *ab initio* experiments. A refinement can be considered successful only if $Q_+$ is large, and then only if the corresponding translation ($\mathbf{y}$ in eq. 23) is small. To evaluate performance, we therefore found the "steepest ascent" maximum of $Q_+$, beginning at $\mathbf{y} = 0$. Successful and failed phase refinements had quite distinct behaviors in $Q_+$ with iteration count, supporting the basin of attraction mechanism. In a successful refinement $Q_+$ rises more or less monotonically, while in the failures $Q_+$ behaved randomly and only rarely exceeded 0.6.

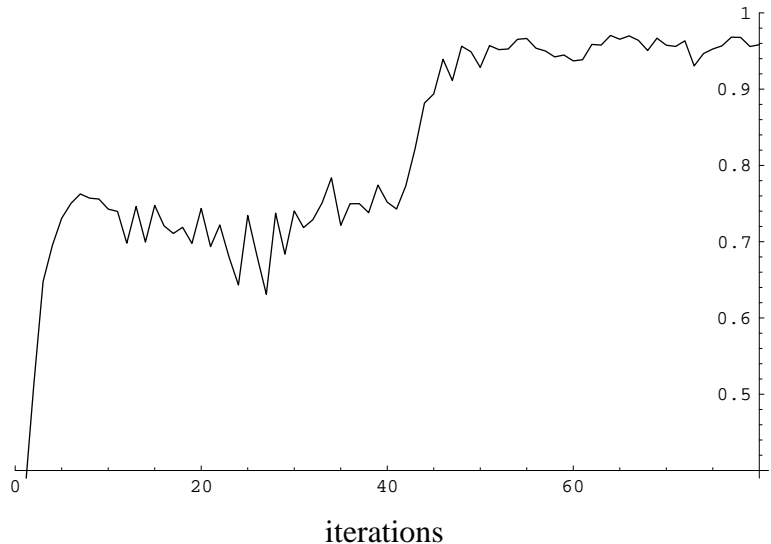

Figure 11. Evolution of the overlap function $Q_+$ for the experiment shown in fig. 10.

In the case of uniform randomization, the success rate makes a rapid transition in a small range of $\delta$ near $\delta = 0.65$. Of ten trials performed at $\delta = 0.6$ all but two yielded successes, while only three out of ten trials at $\delta = 0.7$ gave a successful refinement. With selective randomization, where we vary the number $N_s$, the transition is not quite as abrupt: the success rate for ten trials was 100% at $N_s = 90$, 60% at $N_s = 70$, and 40% at $N_s = 50$. These two experiments give complementary information about the size of the attractive basin: its angular range ($\pm 0.65\pi$) and the extent of its domination by strong reflections (about 60 out of 178).

# 5. Conclusions

The principle of minimum charge exploits the fact that in an *atomic* density the lowest density values — those near zero — are also by far the most common. This lowest density value percolates throughout the unit cell as a distinctive "condensate" that only the correct phase assignment can reproduce. A simple algorithm for finding this singular state proceeds by modifying the phases, iteratively, so as to raise the lowest minimum in the reduced density ( $\tilde{\rho}(\mathbf{x}_{min})$ ). The inevitable outcome of this procedure is a set of local minima that are all nearly degenerate and collectively represent the sought for condensate. As an optimization problem, the quantity being minimized is the minimum average charge density ($\rho_0$) consistent with the requirement of positivity ( $\rho_0 + \tilde{\rho}(\mathbf{x}_{min}) \geq 0$ ). Appropriately normalized ($\rho_0 / \sqrt{I}$), this same quantity provides a useful figure of merit in phase determination.

Because of its "minimal" assumptions about the true nature of atomic densities, the principle of minimum charge is applicable to quasicrystals as well as ordinary crystals. The results of simple experiments with simulated diffraction data, documented here, are sufficiently encouraging to consider tests involving real data. Currently underway is a project to determine the structure of the quasicrystalline AlPdMn icosahedral phase.

# 6. Appendix: Charge minimization and the knapsack problem

Here we show that a special case of a generalization of the charge minimization problem is equivalent to the "knapsack problem" of combinatorial optimization (Nemhauser and Wolsey, 1988).

We consider the case of centrosymmetric densities when by proper choice of origin the reduced density takes the form

$$\tilde{\rho}(\mathbf{x}) = 2 \sum_{\mathbf{q}=1}^{N} A_{\mathbf{q}} s_{\mathbf{q}} \cos(\mathbf{q} \cdot \mathbf{x}), \tag{26}$$

where $s_{\mathbf{q}}$ is now simply a sign. To completely discretize the charge minimization problem we evaluate each Fourier mode on the lattice generated by $\{\mathbf{a}/m, \mathbf{b}/m, \ldots\}$, where $\{\mathbf{a}, \mathbf{b}, \ldots\}$ are the generators of the crystal's Bravais lattice and $m$ is an integer. We let $\mathbf{x}_{\mathbf{p}}$, with $\mathbf{p} = 1, \ldots, M$, be any one of the $M = m^D$ inequivalent points in the $D$-dimensional unit cell, and define an $M \times N$ matrix by

$$C_{\mathbf{pq}} = 2 A_{\mathbf{q}} \cos(\mathbf{q} \cdot \mathbf{x}_{\mathbf{p}}). \tag{27}$$

We note that the columns of $C$ are orthogonal and individually have zero sum (corresponding to orthogonality with the absent $\mathbf{q} = 0$ column). A discretized version of the centrosymmetric charge minimization problem may now be stated in the following language: Find the set of signs $\{s_1, \ldots, s_N\}$ such that the column vector

$$\tilde{\rho}_{\mathbf{p}} = \sum_{\mathbf{q}=1}^{N} C_{\mathbf{pq}} s_{\mathbf{q}}, \tag{28}$$

has the largest possible minimum element. By choosing a suitably large $m$ (so $M = m^D$ is of order $N$ or greater) the discretized version can be an arbitrarily good approximation of the original problem.

As a generalization of the above problem we relax the orthogonality constraint on the columns while maintaining the vanishing of their sums. If there is any hope of an efficient algorithm to solve the charge minimization problem it is conferred by the orthogonality property which we sacrifice in what follows. Indeed, if we now specialize to the case of matrices $C$ with only two rows, we obtain the knapsack problem — a standard "hard" problem in combinatorial optimization.

Since the column sums are zero, for the case of two rows we have

$$C_{1\mathbf{q}} = -C_{2\mathbf{q}} \equiv c_{\mathbf{q}}, \tag{29}$$

and by redefining the signs $s_{\mathbf{q}}$ we can assume without loss of generality that $c_{\mathbf{q}} > 0$ for all $\mathbf{q}$. We then have

$$\tilde{\rho}_1 = \sum_{\mathbf{q}=1}^{N} c_{\mathbf{q}} s_{\mathbf{q}} = -\tilde{\rho}_2. \tag{30}$$

Again, without loss of generality, we may assume $\tilde{\rho}_1 \leq 0$ since we are free to reverse all the signs. Our optimization problem is now that of finding signs which maximizes the non-positive sum (30). In terms of the variables $\theta_{\mathbf{q}} = (1 + s_{\mathbf{q}})/2$ which take on the values 0 or 1, (30) becomes

$$\tilde{\rho}_1 = \sum_{\mathbf{q}=1}^{N} (2c_{\mathbf{q}}) \theta_{\mathbf{q}} - \sum_{\mathbf{q}=1}^{N} c_{\mathbf{q}} \leq 0. \tag{31}$$

This shows the equivalence with the problem of finding the optimal subset of the positive real numbers $2c_{\mathbf{q}}$ that gives the maximal filling of a "knapsack" having a prescribed size (given as the sum of the $c_{\mathbf{q}}$).